\documentclass[12pt]{iopart}
\usepackage{graphicx}

\begin{document}

\newcommand{\NaCo}{Na$_{x}$CoO$_{2}\cdot y$H$_{2}$O}

\title{Possible spin-triplet superconductivity in {\NaCo} -- $^{59}$Co NMR study --}

\author{M. Kato\dag\footnote[2]{Present address: Department of Molecular Science and Technology, 
Faculty of Engineering, Doshisha University, Kyotanabe, Kyoto 610-0394, Japan.}, C. Michioka\dag, T. Waki\dag, Y. Itoh\dag, K. Yoshimura\dag, K. Ishida\S, H. Sakurai$\|$, E. Takayama-Muromachi$\|$, K. Takada\P\ and T. Sasaki\P}

\address{\dag\ %
Department of Chemistry, Graduate School of 
Science, Kyoto University, Kyoto 606-8502, Japan
}
\address{\S\ %
Department of Physics, Graduate School of 
Science, Kyoto University, Kyoto 606-8502, Japan
}
\address{$\|$\ %
Superconducting Materials Center, National Institute for Materials Science, 1-1 Namiki, Tsukuba, Ibaraki 305-0044, Japan
}
\address{\P\ %
Advanced Materials Laboratory, National Institute
for Materials Science, Namiki 1-1, Tsukuba, Ibaraki 305-0044, Japan
}

\ead{makato@mail.doshisha.ac.jp}

\begin{abstract}
We report $^{59}$Co NMR studies on the magnetically oriented powder samples of Co-oxide superconductors {\NaCo} with $T\mathrm{_{c} \sim 4.7 K}$.
From two-dimensional powder pattern in the NMR spectrum, the $ab$-plane Knight shift in the normal state was estimated by 
the magnetic field dependence of second-order quadrupole shifts at various temperatures. 
Below 50 K, the Knight shift shows a Curie-Weiss-like temperature dependence, similarly to the bulk magnetic susceptibility $\chi$. From the analysis of so-called $K$-$\chi$ plot, the spin and the orbital components of $K$ and the positive hyperfine coupling constant were estimated. 
The onset temperature of superconducting transition in the Knight shift does not change so much in an applied magnetic field up to 7 T, which is consistent with the reported high upper critical field $H\mathrm{_{c2}}$.
The Knight shift at 7 T shows an invariant behavior below $T\mathrm{_{c}}$. 
No coherence peak just below $T\mathrm{_{c}}$ was observed in the temperature dependence of the nuclear spin-lattice relaxation rate $1/T_1$ in both cases of NMR and NQR. We conclude that the invariant behavior of the Knight shift below $T\mathrm{_{c}}$ and unconventional behaviors of $1/T_1$ possibly indicate the spin-triplet superconductivity with $p$- or $f$-wave symmetry.
\end{abstract}

\submitto{\JPCM}


\maketitle

\section{Introduction}

\hspace*{10mm}Recent discovery of superconductivity in {\NaCo} with the transition temperature ($T\mathrm{_{c}}$) of about 4.7 K \cite{Takada} has been a major breakthrough in search for novel layered transition metal oxide superconductors. 
{\NaCo} has a two-dimensional (2D) crystal structure where Co atoms form a 2D triangular lattice separated by Na$^{+}$ ions and H$_2$O molecules. 
The intercalation of H$_2$O molecules leads to an increase of the $c$-axis lattice constant, which is considered to enhance two dimensionality of CoO$_2$ layers. 
The triangular configuration on the $\mathrm{CoO_{2}}$ plane potentially  has a possibility of an unconventional superconductivity by marvelous mechanism in the novel quantum state, as suggested by many theoretical works. 
For example, Tanaka $et$ $al$. indicated that the spin-triplet superconductivity may be realized with similar pairing process to $\mathrm{Sr_{2}RuO_{4}}$ since the hexagonal structure in {\NaCo} is favorite for a spin-triplet pairing \cite{Tanaka}. 
Some groups studied a single band $t$-$J$ model as a prototype of strong coupling electrons theories to understand the low energy electronic phenomena based on the resonating-valence-bond (RVB) state \cite{Anderson, baskaran1, kumar, wang, ogata}. 
Kuroki $et$ $al$. introduced a single band effective model with taking into account pocket-like Fermi surfaces along with van Hove singularity near the K point  in the first Brillouin zone \cite{kuroki1, kuroki2, kuroki3, kuroki4}. 
They showed that the large density of states near the Fermi level gives rise to ferromagnetic spin fluctuations, leading to the $f$-wave superconductivity due to  disconnected Fermi surfaces near the $\Gamma$ point. 
Khaliullin $et$ $al.$ emphasized the importance of a spin-orbit (LS) coupling, and showed that the hole carriers with pseudo-spin one-half on a $f$ level, separated from $t_{2g}$-orbital degeneracy due to trigonal distribution and spin-orbit interaction, makes the spin-triplet state always be favored \cite{khaliullin}.

In order to clarify the superconducting mechanism in {\NaCo}, it is very important to investigate the Cooper-pair symmetry. 
Nuclear magnetic resonance (NMR) is known as a powerful probe to investigate the pairing symmetry in the superconducting state. 
In this paper, we report the detailed experimental results of Knight shift, $K$, and the spin-lattice relaxation rate, $1/T_1$, in {\NaCo} obtained by $^{59}$Co NMR and  nuclear quadrupole resonance (NQR) measurements. 
We demonstrate that $K_{y}$, one of the in-plane Knight shifts is invariant below $T\mathrm{_{c}}$ at a high field of about 7 T, where $1/T_1$ drops at $T\mathrm{_{c}}$ and obeys $T^{3}$ law without the coherence peak just below $T\mathrm{_{c}}$. These results are attributable to the possible formation of $p$- or $f$-wave superconducting state with line-nodes on the superconducting gap. 
Our results are inconsistent with the previous report by Kobayashi $et$ $al$., which suggested spin-singlet $s$-wave superconductivity from the result of a lower field of about 2 T \cite{Kobayashi}. 
However, the precise values of the Knight shift cannot be estimated without explaining the frequency dependence of field-swept spectra because of the coexistence of electric quadrupole and Zeeman interactions.   
Thus, we clearly present how to estimate the spin part of the Knight shift using the frequency dependence of resonance fields above $T\mathrm{_{c}}$. 
The full analysis of the NMR parameters is crucially important for discussing the Cooper-pair symmetry in the superconducting state. 

\section{Experiments}

\hspace*{10mm}The powder sample of {\NaCo} 
was prepared by the oxidation process from the mother compound, $\mathrm{Na_{0.7}CoO_{2}}$ \cite{Takada}. 
The obtained sample was confirmed to be a single phase by the powder X-ray diffraction (XRD) measurement. 
The chemical compositions of Na and H$_2$O were found to be about 0.35 and 1.3, respectively, by inductively coupled plasma atomic emission spectroscopy (ICP) \cite{Takada}. 
The superconducting transition temperature ($T\mathrm{_{c}}$) was estimated from the temperature dependence of the magnetization using SQUID magnetometer. 
$\mathrm{^{59}}$Co NMR and NQR measurements were performed by the spin-echo method with a standard phase coherent-type pulsed spectrometer. 
The powder sample was oriented and fixed by using an organic solvent hexane under the condition of the external field, $H$ = 8 T. 
In our NMR measurement, hexane was found to be an optimal agent to fix the sample because it has the melting point of 178 K and does not absorb water molecules. 
Stycast 1266, which is useful for high $T\mathrm{_{c}}$ cuprates superconductors, were not suitable owing to their desiccating effect. 
From the XRD measurement, we found that the $c$-axis of the sample was oriented perpendicular to the external magnetic field $H$ and that the random orientation in the $c$-plane was obtained along $H$. 
Since the orientation and the NMR measurements were performed at the same time, the magnetic field in NMR was applied completely to the same direction where the sample was aligned.
NQR measurements were performed in zero field. 
$1/T_{1}$ was estimated by monitoring the recovery of the spin-echo amplitude, $M(t)$, after an inversion pulse as a function of the delayed time ($t$) between an inversion $\pi$ and the observation radio-frequency $\pi/2-\pi$ pulses. 

 Once the sample was cooled, it was kept below 100 K even during the intermission of NMR and NQR measurements.
 We must maintain a degree of the orientation throughout the measurements because the resonance fields for peaks and shoulders of the NMR spectrum are sensitive to the sample orientation. 
 We also notice that it is extremely important to avoid the deficiency of H$_2$O molecules since the superconducting properties such as $T\mathrm{_{c}}$ are sensitive to the H$_2$O content \cite{Ohta}. 

\section{Results and discussion}
\subsection{$\mathrm{^{59}Co}$ Knight shift above $T\mathrm{_{c}}$}
\hspace*{10mm}Figure 1 shows a typical $\mathrm{^{59}Co}$ NMR spectrum of {\NaCo} at a frequency of 59.200 MHz.
\begin{figure}[h]
 \begin{center}
 \includegraphics[width=0.8\linewidth]{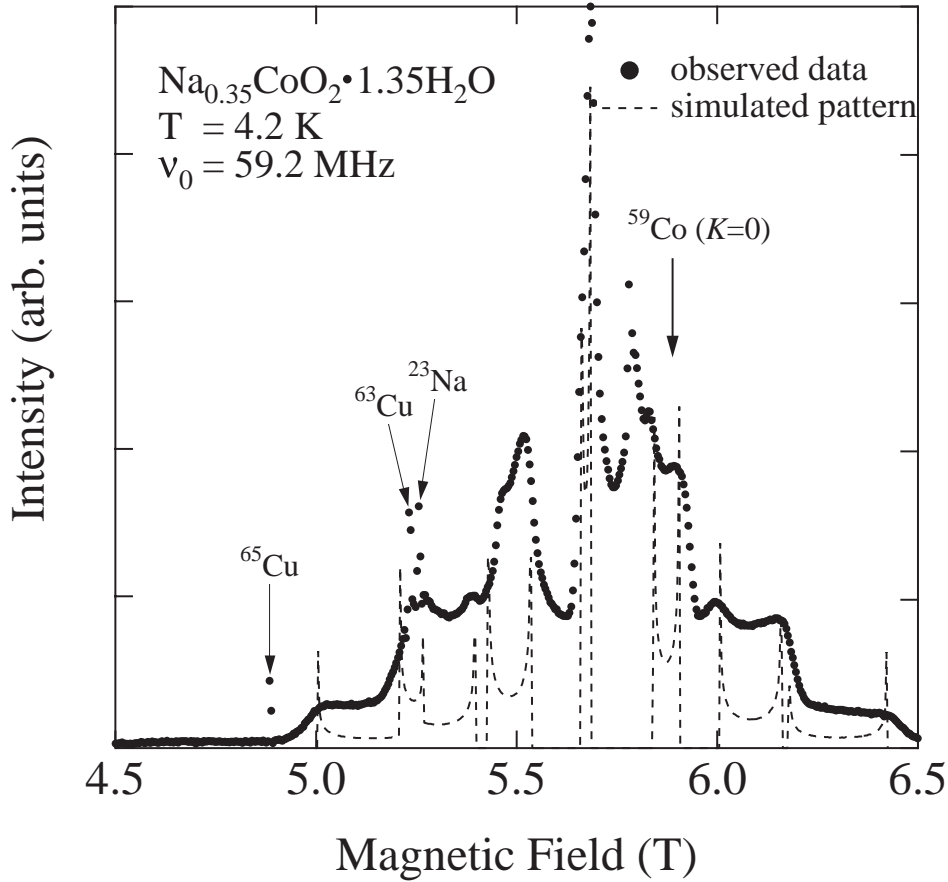}
 \end{center}
 \caption{\label{fig:Spectrum4KNS}
Field-swept NMR spectrum of {\NaCo} at a constant frequency ($\nu_0$) of 59.2 MHz taken at 4.2 K. Arrows indicate the resonance fields of $^{63,65}$Cu nuclei in the resonance coil, $^{23}$Na nuclei and the Knight shift reference position ($K$ = 0) of $^{59}$Co nuclei. Filled circles show the observed data. Dashed line indicates the peak positions and intensities simulated for the case of an anisotropic Knight shift and a second order quadrupole effect with $K_{x}$ = 3.56 \%, $K_{y}$ = 3.21 \%, $\nu _Q$ = 4.15 MHz and the asymmetric parameter of electric field gradient $\eta$ = 0.21. }
\end{figure}
The observed NMR spectrum can be understood by a 2D powder pattern with anisotropic Knight shifts and the second order quadrupole interaction. 
We have succeeded in determining these NMR parameters from the following analysis. 

\begin{figure}
\begin{center}
\includegraphics[width=0.47\linewidth]{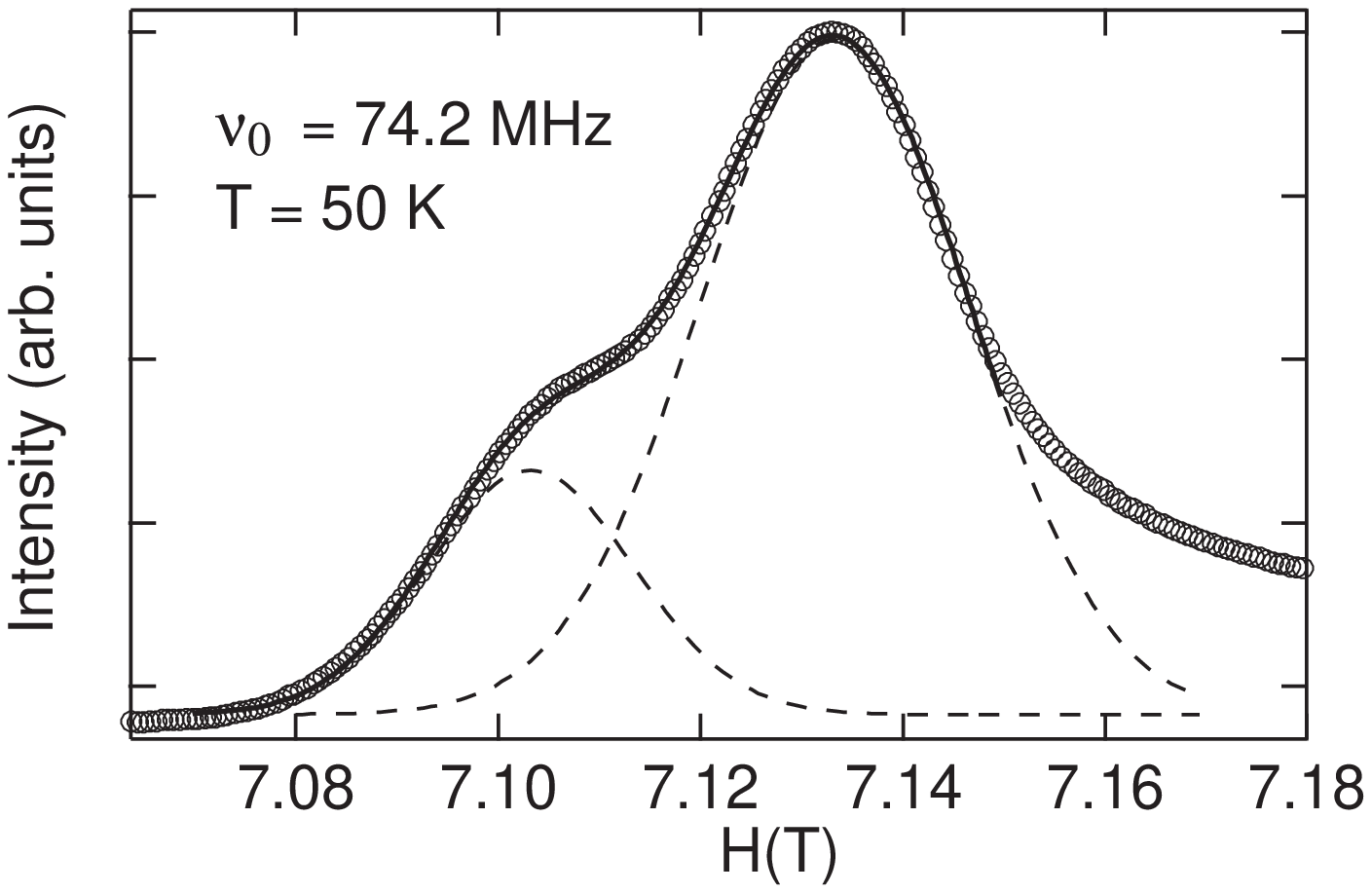}
\end{center}
\caption{\label{fig:peak} Field-swept NMR spectrum of $\mathrm{Na_{0.35}CoO_{2} \cdot 1.3H_{2}O}$ at $T$ = 50 K and $\nu_0$ = 74.2 MHz. Two peaks, which are decomposed by the Gaussian fitting indicated by the dashed lines, correspond to the central resonance of the $I_{z} = -1/2 \leftrightarrow 1/2$ transition. The anisotropic Knight shifts $K_{x}$ and $K_{y}$ were estimated from the low-field smaller peak and the high-field larger peak, respectively (see the text). }
\end{figure}

In order to determine the Knight shift above $T\mathrm{_{c}}$, we use the central transition line ($+1/2 \leftrightarrow -1/2$) as shown in Fig. \ref{fig:peak}.  
We assume that principal axes of the Knight shift tensor coincide with those of the electric-field gradient (EFG) tensor. 
This assumption is generally rational because those tensors are mainly governed by the local symmetry around the Co site. 
Because of the quadrupole effect up to the second-order perturbation, a conventional 3D powder pattern of the central transition has five specific resonance frequencies $\nu_1$-$\nu_5$ in a constant field which are equivalent to resonance fields $H_1$-$H_5$ in a constant frequency ($\nu_0$) \cite{text}.
These fields can be observed as two peaks ($H_2$ and $H_4$), two shoulders ($H_1$ and $H_5$) and a step ($H_3$) in the case of $\eta < \frac{1}{3}$, where $\eta$ is the asymmetric parameter of the EFG tensor defined by $\eta = |V_{XX}-V_{YY}|/|V_{ZZ}|$, in which $V_{ii}$ ($ii$ = $XX$, $YY$, $ZZ$) are the principal components of the EFG tensor with the relation, $|V_{XX}| \le |V_{YY}| \le |V_{ZZ}|$ \cite{text}. 
In the present case of the 2D powder spectrum, only two peaks appear at specific resonance fields ($H\mathrm{_{1}}$ and $H\mathrm{_{2}}$) because of the sample orientation. 
Hence the central spectrum should consist of two peaks at $H_1$ and $H_2$. 
Actually, the central resonance was decomposed into two Gaussian components using the least-square fitting to determine the experimental values of $H_1$ and $H_2$ as shown in Fig. \ref{fig:peak}. 
%
\begin{figure}
\begin{center}
\includegraphics[width=0.8\linewidth]{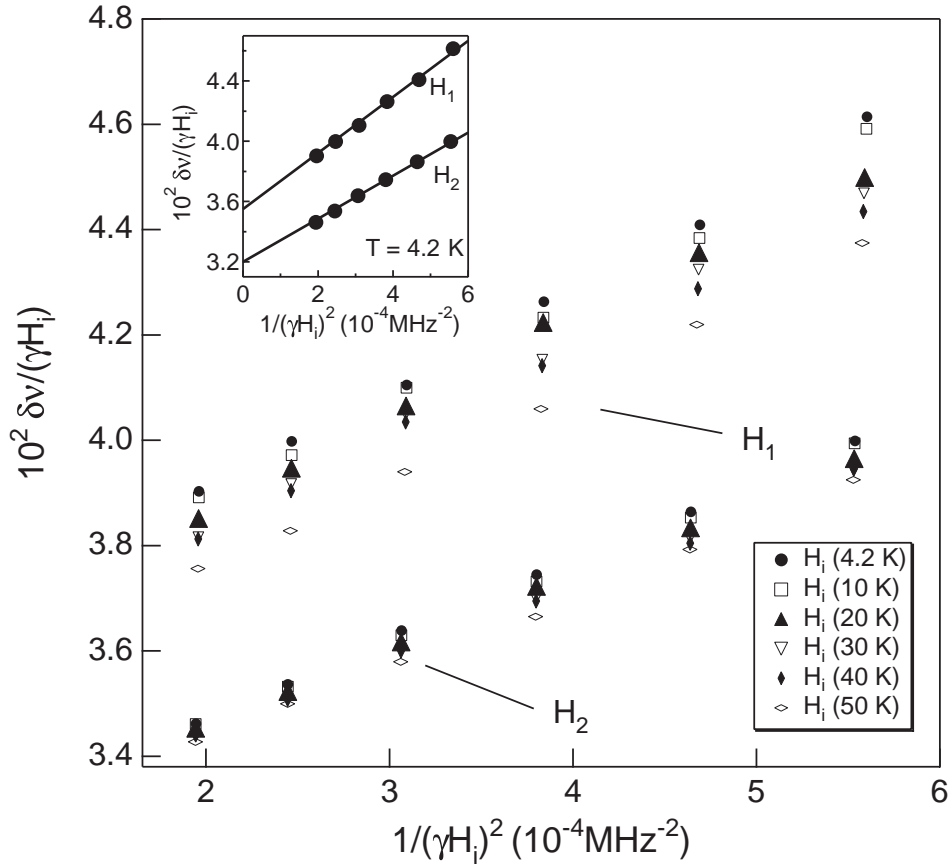}
\end{center}
\caption{\label{fig:aboveTc} Second-order quadrupole plot at various temperatures above $T\mathrm{_{c}}$ and at 4.2 K (just below $T\mathrm{_{c}}$) obtained at $\nu_0$ ranging from 44.2 to 74.2 MHz. $H\mathrm{_{1}}$ and $H\mathrm{_{2}}$ are the resonance fields at which the $\mathrm{^{59}Co}$ NMR spectrum with 2D powder pattern of the central transition show two singular points. Inset shows the data at 4.2 K with the fitted lines using the NMR parameters, $\nu_{Q}$ = 4.15 MHz, $\eta$ = 0.21, $K_{x}$ = 3.56 \% and $K_{y}$ = 3.21 \%. }
\end{figure}
These resonance fields at a constant frequency $\nu_0$ are expressed by $K_{j}$, principal values of the Knight shift tensor, and $\nu_Q$, the quadrupole frequency, as follows,  
\begin{eqnarray}
\label{eq:A} \frac{\delta\nu_{i}}{\gamma\mathrm{_{N}}H_{i}} = K_{j} + \frac{C_{i}}{(1+K_{j})(\gamma\mathrm{_{N}}H_{i})^{2}}, \\
\label{eq:B} \delta\nu = \nu_{0} - \gamma_\mathrm{{N}}H_{i}, \\
\label{eq:C} C_{1} = \frac{R(3+\eta)^{2}}{144},  C_{2} = \frac{R(3-\eta)^{2}}{144}, \\
\label{eq:D} R = \nu_{Q}^{2}[I(I+1)-\frac{3}{4}],
\end{eqnarray}
where the values with $j = X, Y$ correspond to those with $i$ = 1, 2, respectively, $\gamma\mathrm{_{N}}$ is the nuclear gyromagnetic ratio (10.054 MHz/T for $\mathrm{^{59}Co}$ with the nuclear spin of $I$ = 7/2) \cite{text}. 
As seen in Eq. (\ref{eq:A}), the linear relation should be expected when $\delta\nu_{i}$/$\gamma\mathrm{_{N}}H_{i}$ is plotted against $(\gamma\mathrm{_{N}}H_{i})^{-2}$. 
Then, the intercepts on the vertical axis at $(\gamma H_{i})^{-2} \to 0$ give the values of $K_x$ and $K_y$ independent of $\nu_Q$. 

Figure \ref{fig:aboveTc} shows the experimental values of  $\delta\nu/(\gamma\mathrm{_{N}}H_{i})$ plotted against $(\gamma\mathrm{_{N}}H_{i})^{-2}$ ($i$ = 1, 2) measured 
above $T\mathrm{_{c}}$ and at 4.2 K in the frequency range of 44.2 - 74.2 MHz. 
This plot is referred to a second-order quadrupole plot. 
The experimental values of $\delta\nu_i / \gamma H_i$ clearly lie on the respective straight lines against $(\gamma H_j)^{-2}$ in the plots. 
One should note that $K_{x}$ and $K_{y}$ (interceptions of the $\delta\nu_{i}$/$\gamma\mathrm{_{N}}H_{i}$ axis) decrease with heating while the slopes are almost independent of $T$. 
Since the slope of the line, $C_{i}/(1+K_j) (\simeq C_i)$, is determined by $\nu_Q$ and $\eta$, the almost invariant slope indicates that $\nu_Q$ and $\eta$ are almost constant in this temperature range. 
This result is consistent with our NQR experiment (not shown here). 
The inset in Fig. \ref{fig:aboveTc} shows the result of the least-square fitting using Eqs. (\ref{eq:A})-(\ref{eq:D}) at 4.2 K. 
The obtained NMR parameters are $K_{x}=3.56 \%$, $K_{y}=3.21 \%$, $\nu_{Q}$ = 4.15MHz and $\eta$ = 0.21. 
These parameters at another temperatures were estimated similarly. 
Using these parameters, we can calculate the simulated spectra as shown by dashed lines in Fig. \ref{fig:Spectrum4KNS}. 
Consequently, the simulated spectrum including first, second and third satellites is in good agreement with the observed one. 

Within the analysis mentioned above, we could not estimate the $Z$ component of the Knight shift due to the 2D sample orientation. 
However, we have estimated this value as follows. 
In Fig. \ref{fig:Spectrum4KNS}, we noticed the peak around $H$ = 5.8 T, which is not usually expected for the 2D pattern. We regarded this peak as the NMR signals from the accidentally aligned powders along the $c$-axis. 
Assuming that this peak corresponds to $H_3$, another step in the powder spectrum (a peak in the $c$-axis oriented spectrum), we have estimated the value of $K_z$ = 1.88 \%. 

\begin{figure}
\begin{center}
\includegraphics[width=0.8\linewidth]{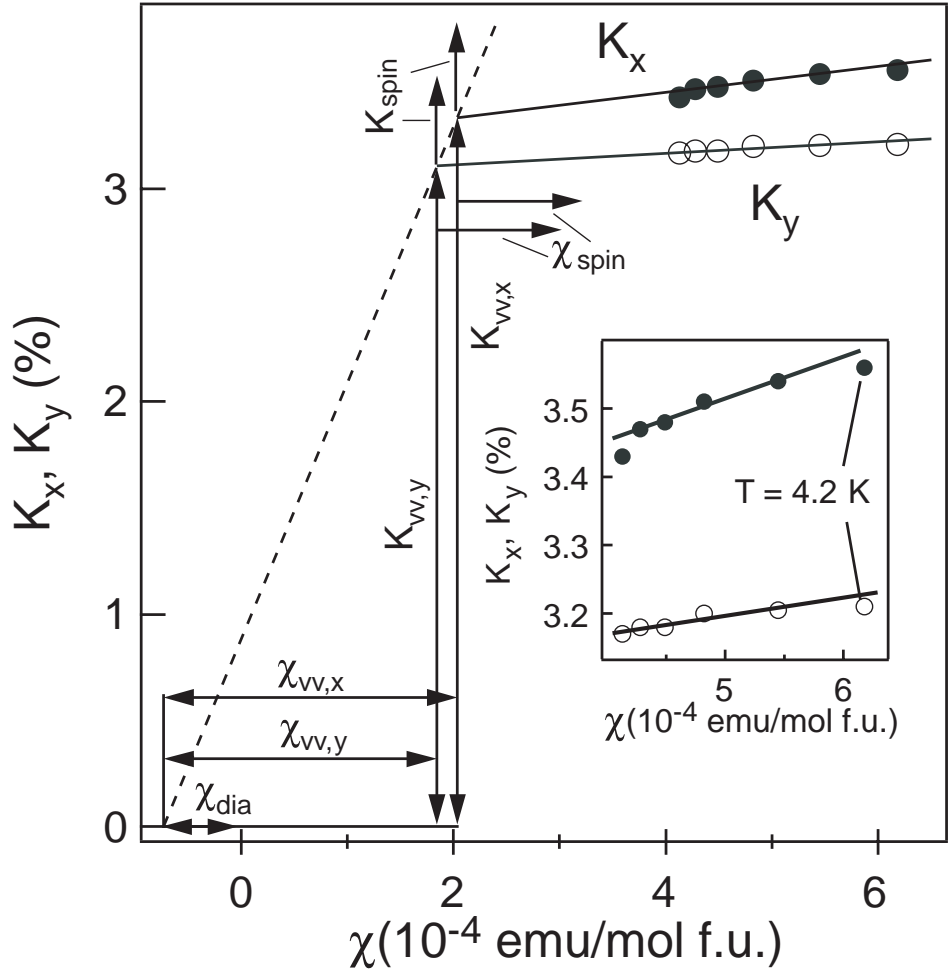}
\end{center}
\caption{\label{fig:Kchi} Knight shift vs. susceptibility diagram in $\mathrm{Na_{0.35}CoO_{2} \cdot 1.3H_{2}O}$. The data at 4.2 K (below $T\mathrm{_{c}}$) is shown for comparison but not used for the line fitting. Inset shows the expansion. The numerical details for constructing the diagram are given in the text. The dashed line is the orbital shift against the orbital susceptibility, $K_{VV}= A_{hf}^{VV} \times \chi_{VV}$.}
\end{figure}

In Fig. \ref{fig:Kchi}, we indicate the obtained $K_x$ and $K_y$ against the uniform magnetic susceptibility with temperature as an implicit parameter. This is so-called $K$-$\chi$ plot, which is useful for the analysis of various contribution to the Knight shift \cite{Jaccarino}.
For the 3$d$-transition oxides, the Knight shift can be expressed by
\begin{equation}
K(T) = K_{spin}(T) + K_{s} + K_{VV},
\label{eq:E}
\end{equation}
where $K_{spin}(T)$, $K_{s}$ and $K_{VV}$ are a $T$-dependent spin shift, a $T$-independent spin shift and a Van Vleck orbital shift, respectively.
The magnetic susceptibility can be expressed by,
\begin{equation}
\chi(T) = \chi_{spin}(T) + \chi_{s} + \chi_{VV}+\chi_{dia},
\label{eq:chi}
\end{equation}
where $\chi_{spin}(T)$, $\chi_{s}$, $\chi_{VV}$ and $\chi_{dia}$ are a $T$-dependent spin susceptibility, a $T$-independent spin susceptibility, a Van Vleck orbital susceptibility and a diamagnetic susceptibility at core electrons, respectively. 
The hyperfine coupling constants connects the respective components of the $^{59}$Co Knight shift in Eq. (\ref{eq:E}) to those of the magnetic susceptibilities in Eq. (\ref{eq:chi}) as follows,
\begin{eqnarray}
K_{spin}(T) &=& A_{hf} \chi_{spin}(T), \\
K_{s} &=& A_{s} \chi_{s}, \\
K_{VV} &=& A_{VV} \chi_{VV}.
\label{eq:coupling}
\end{eqnarray}

The diamagnetic susceptibilities of core electrons can be estimated from relativistic Hartree-Fock calculations. From $\chi_{dia}^\mathrm{{H}}$ = $ - 0.4 \times 10^{-9}$ $\mu\mathrm{_{B}/atom}$, $\chi_{dia}^\mathrm{{O}}$ = $ - 1.6 \times 10^{-9}$ $\mu\mathrm{_{B}/atom}$, $\chi_{dia}^\mathrm{{Na}}$ = $ - 3.8 \times 10^{-9}$ $\mu\mathrm{_{B}/atom}$ and $\chi_{dia}^\mathrm{{Co}}$ = $ - 5.5 \times 10^{-9}$ $\mu\mathrm{_{B}/atom}$ \cite{Mendelsohn}, 
the total diamagnetic susceptibility of {\NaCo} was estimated to be $\chi_{dia} = \mathrm{7.3\times10^{-5} emu/mol}$. 
The hyperfine coupling constant for the $d$ orbital Van Vleck component, $A_{hf}^{VV} = K_{VV}/\chi_{VV} = 6.7\times10^{2}$ kOe/$\mu_{B}$ was assumed using $2/\langle r^{-3} \rangle = 9.05\times10^{25}\mathrm{cm}^{-3}$ for $\mathrm{Co^{3+}}$ and the reduction factor of 0.8 for the metallic state in compared with the free ion. 
Using these parameters, the coupling constants for the temperature dependent Knight shift components, $A_{hf}^{x}$ and $A_{hf}^{y}$ were estimated to be +34 kOe/$\mu_{B}$ and +14.7 kOe/$\mu_{B}$ by the least-square fitting shown with solid lines in Fig. \ref{fig:Kchi}. 
Then, $\chi_{VV,x}$ and $\chi_{VV,y}$ were estimated to be $2.0\times10^{-4}$ and $1.8\times10^{-4}$ emu/mol f.u., respectively, and $K_{VV,x}$ and $K_{VV,y}$ to be 3.3 and 3.1 $\%$, respectively, at most. Therefore the spin Knight shifts were found to be $\sim 0.3 \%$ for $K_{x}$ and $\sim 0.1 \%$ for $K_{y}$ at least. 

The hyperfine coupling constant $A_{hf}$ between the Co nuclear spin and the electron spins can be expressed by
\begin{equation}
A_{hf} = A_{cp} + A_{dip} +A_{orb} +A_{tr},
\label{eq:hf}
\end{equation}
where the on-site hyperfine coupling constants of $A_{cp}$, $A_{dip}$ and $A_{orb}$ are originated from core polarization, dipolar interaction, spin-orbit coupling, respectively, and $A_{tr}$ is from the supertransferred hyperfine field. In 3$d$ transition metals, the hyperfine coupling constant due to the core spin polarization of 3$d$ electrons is $negative$ with $A_{cp} \approx -100$ kOe$/\mu_{B}$ \cite{Freeman}. The observed $positive$ hyperfine coupling constant of $\mathrm{Na_{0.35}CoO_{2} \cdot 1.3H_{2}O}$ is unusual. For some Co-oxides \cite{Mitoh,Miyatani,Hayakawa} and alloys \cite{Narath,Dupree}, the $positive$ hyperfine coupling constant of $^{59}$Co nuclei was observed and attributed to the presence of a large LS-coupling effect. 
In the present case, a large orbital contribution to $K_{spin}$ via the LS-coupling can account for the $positive$ $A_{hf}$.

In our preliminary experiment for an nonoriented 3D powder sample, $K_{z}$ is also estimated to be positive $\sim 1.88\%$. 
Probably the positive coupling constant is caused by not only the anisotropic $A_{orb}$ and $A_{dip}$ but also the transferred hyperfine field $A_{tr}$. 
In the present stage, we cannot estimate separately $A_{orb}$ and $A_{tr}$. The value of $K_{y}$ is consistent with that reported by Mukhamedshin $et$ $al. $\cite{mukhamedshin}.

At first, in the analysis in Fig. \ref{fig:Kchi}, we assumed all the temperature independent Knight shift as $K_{VV}$. However, we must consider the possibility that a part of the spin term does not depend on temperature and $K_{VV}$ is smaller than the estimated value, which should be important when two bands are dominant in the Fermi surface. That is, there must be a finite contribution of $K_{s}$ in Eq. (\ref{eq:E}) and $\chi_{s}$ in Eq. (\ref{eq:chi}).

\subsection{$\mathrm{^{59}Co}$ Knight shift below $T\mathrm{_{c}}$}

\begin{figure}[h]
 \begin{center}
 \includegraphics[width=0.5\linewidth]{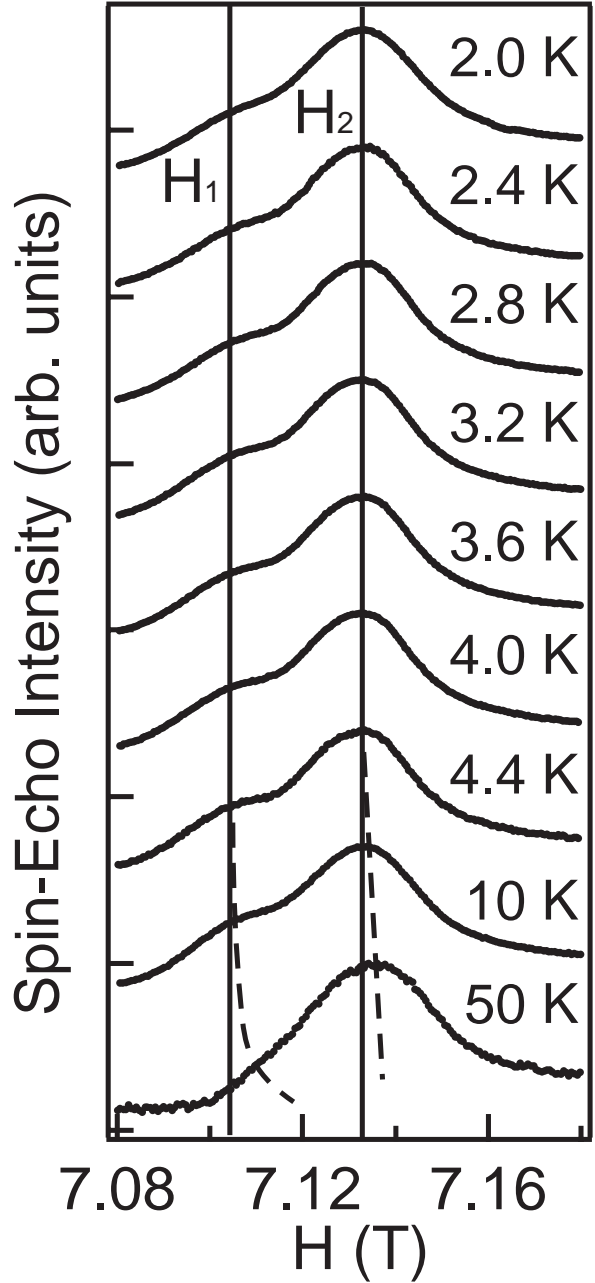}
 \end{center}
 \caption{\label{fig:center_line}
Temperature dependence of the central resonance in the spin echo spectra at $\nu_0$ of 74.2 MHz. Solid lines and dashed curves are guides for eyes. The central resonance field does not change below $T\mathrm{_c}$.}
 \end{figure}

\hspace*{10mm} Figure \ref{fig:center_line} shows the temperature dependence of the central spectra at higher field (about 7 T).
Since the estimated upper critical field $H\mathrm{_{c2}}$ of {\NaCo} was reported to be high, 61 T by the magnetization measurement \cite{Sakurai} and 17.1 T by the heat capacity measurement \cite{Jin}, it should be noted that $T\mathrm{_{c}}$ is almost constant, and that samples were in the superconducting state below $T\mathrm{_{c}}$ in our experimental condition.
As seen in Fig. \ref{fig:center_line}, the central resonance fields $H_1$ and $H_2$ above $T\mathrm{_{c}}$ shift to higher values with increasing temperature, suggesting that the spin contributed Knight shift increases owing to the positive hyperfine coupling constants with a second-order quadrupole shift as discussed above. 
On the other hand, the values of $H_1$ and $H_2$ do not change below $T\mathrm{_{c}}$, indicating that $K_x$ and $K_y$ are almost constant at 3.56 and 3.21 \% below $T\mathrm{_{c}}$ under the field of about 7 T within an experimental precision. 
In the BCS theory, the spin susceptibility, $\chi _{spin}$, in the superconducting state is described as 
 \begin{equation}
 \chi _{spin} = -4\mu _{B}^{2}\int _{0}^{\infty}N_{s}(E)\frac{\partial f(E)}{\partial E}dE,
 \end{equation}
 where $\mu _{B}$ is Bohr magneton, $N _{s}(E)$ is the quasiparticle density of states, and $f(E)$ is the Fermi-Dirac distribution function.
It is widely known for a singlet pairing superconductor that the spin Knight shift decreases with cooling and vanishes at $T=0$ in the Yosida function being proportional to $e^{-\Delta(0)/kT}$ at $T \ll T\mathrm{_{c}}$ \cite{Yosida}. 
The mean free path $l$ is estimated to be $l \geq 100$ {\AA} from the low-temperature resistivity \cite{Chou} and the Fermi wave-number $k_\mathrm{F}$ \cite{Yang}, and the coherence length $\xi$ to be $\xi \sim$ 23.2 \AA \cite{Sakurai}. 
Since the ratio of $l/\xi$ is larger than 4, this system can be considered to be a clean superconductor. 
Thus, the invariant behavior of $K_x$ and $K_y$ at about 7 T suggests a spin-triplet $p$- or $f$-wave pairing superconductivity.

\begin{figure}
\begin{center}
\includegraphics[width=0.68\linewidth]{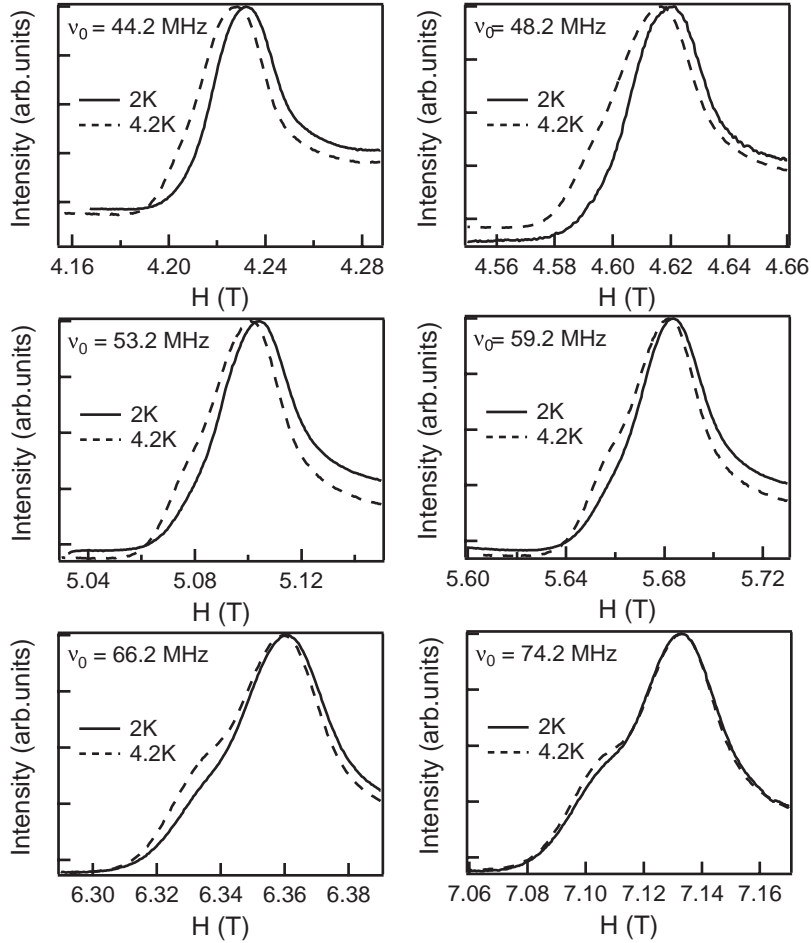}
\end{center}
\caption{\label{fig:spectraLF} Field-swept 2D powder spectra for the central transition at 2.0 and 4.2 K taken at several fixed frequencies $\nu_0$.}
\end{figure}

Figure \ref{fig:spectraLF} shows the central peaks in the field-swept spectra at 2.0 and 4.2 K at several constant frequencies (44.2 - 74.2 MHz). 
%
%
\begin{figure}
\begin{center}
\includegraphics[width=0.8\linewidth]{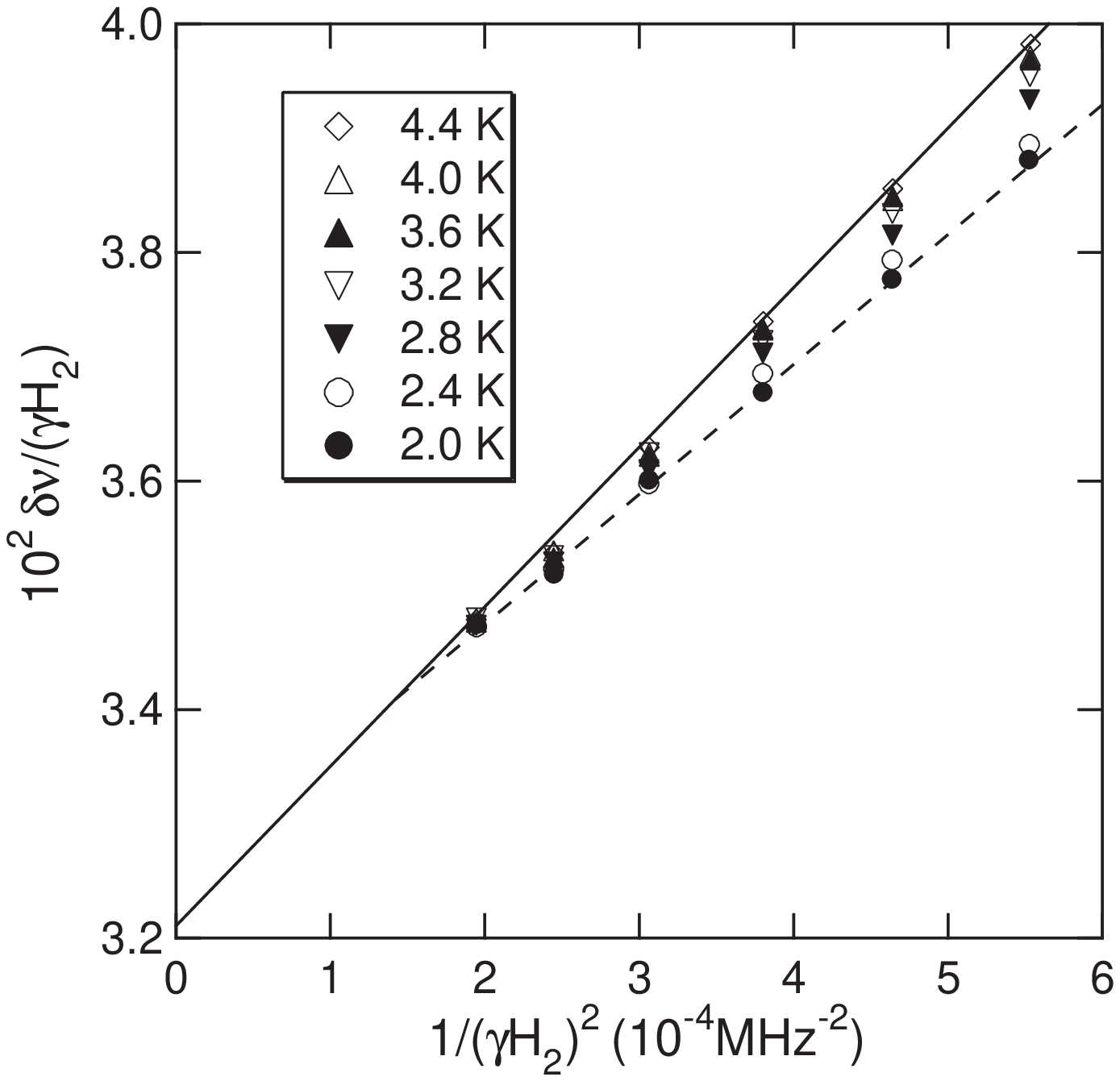}
\end{center}
\caption{\label{fig:belowTc} Second-order quadrupole plots for $H_2$ at various temperatures below $T\mathrm{_{c}}$. Solid line is the same one for $H_2$ as shown in the inset of Fig. \ref{fig:aboveTc}. Dashed line is a guide for eyes. 
The slope of the linear correlation seems to change below $T\mathrm{_{c}}$. 
This is not due to the change of $\nu_{Q}$ nor $\eta$ but maybe due to the diamagnetic effect.}
\end{figure}
The resonance fields are almost the same between 2.0 and 4.2 K at $\nu_0$ = 74.2 MHz, but it changes drastically at $\nu_0$ = 44.2 MHz. 
The difference of the resonance fields between 2.0 and 4.2 K gradually increases with decreasing the experimental frequency. 
As well as in the normal state, we discuss the second-order quadrupole plot in the superconducting state shown in Fig. \ref{fig:belowTc}. 
Only $H_2$ was estimated from the peaks in the spectra, because it was difficult to decompose the spectra especially at lower frequencies. 
In Fig. \ref{fig:belowTc}, the experimental values at 4.4 K show a linear relation similar to those above $T\mathrm{_{c}}$ in Fig. \ref{fig:aboveTc}. 
On the other hand, the slope in the superconducting state seems to decrease with temperature, that is, a non-linear relation appears below $T\mathrm{_{c}}$. 
Since $\nu_{Q}$ and $\eta$ are almost constant also below $T\mathrm{_{c}}$, which were confirmed by our NQR measurements, the decrease of the slope is not due to the change in the quadrupole shift below $T\mathrm{_{c}}$. Thus, the Knight shift itself must depend on the applied magnetic field.

\begin{figure}[h]
\begin{center}
\includegraphics[width=0.8\linewidth]{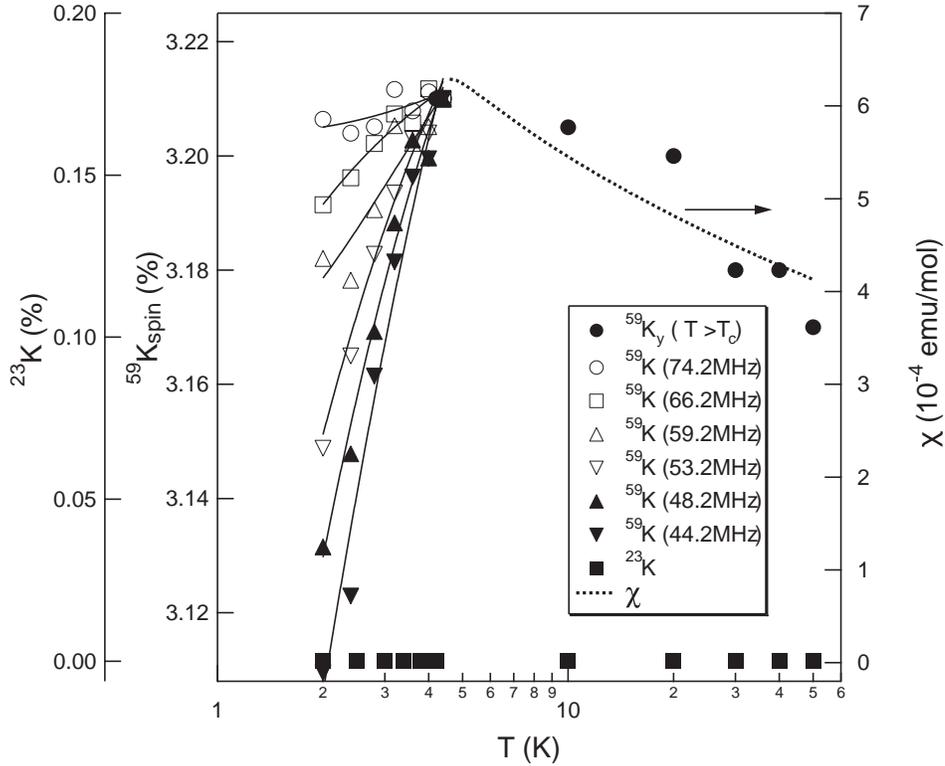}
\end{center}
\caption{\label{fig:KTbelowTc} 
Temperature dependence of the estimated spin part of the Knight shift at various resonance frequencies $\nu_0$. The data of $K_{y}$ below $T_{\mathrm{c}}$ were obtained by extrapolation from the respective data points in Fig. \ref{fig:belowTc} and Eq. (\ref{eq:A}) assuming that $\nu_Q$ and $\eta$, then $C_{i}$ in Eq. (\ref{eq:A}), are the same as those at 4.2 K. Here, a dashed line indicates the temperature dependence of $\chi$. Solid lines are guides for eyes.
}
\end{figure}

In Fig. \ref{fig:KTbelowTc}, we estimate the temperature dependence of the spin part of the $^{59}$Co Knight shift at various resonance frequencies $\nu_0$ below $T_{\mathrm{c}}$ and the Knight shift of $\mathrm{^{23}Na}$ nuclei. This estimation was done by extrapolating from the respective data points in the second-order quadrupole plot in Fig. \ref{fig:belowTc} and Eq. (\ref{eq:A}), assuming that all the slopes, $C_{i}$ in Eq. (\ref{eq:A}), were fixed to be the same value at 4.2 K.
As shown in Fig. \ref{fig:KTbelowTc}, the temperature-dependent term of the shift increases with decreasing $\nu_0$. 
One should note that the values of $T_{\mathrm{c}}$ onsets are almost invariant with changing the magnetic field, which is consistent with the rather high value of $H_{c2}$. 
The behavior of the Knight shift below $T\mathrm{_{c}}$ can be explained by a superconducting diamagnetic shift or a field-induced spin Knight shift. 

At first we discuss the possibility of the site-dependent diamagnetic effect.
The Knight shift of $\mathrm{^{23}Na}$ does not change with $K\sim0$ indicating that the spin contribution is small and the diamagnetic shift in the superconducting state is also small. The effect of diamagnetic shift was remarkable at the $\mathrm{^{59}Co}$ nuclear sites but negligible at the $\mathrm{^{23}Na}$ sites. If Josephson vortices penetrate into the Na layers, the diamagnetic shift due to supercurrents would be expected to be large at the Co sites but not at the Na site. 
Theoretically, in the vortex state with extremely strong two-dimensional system, that is, with the short coherence length along $c$-axis, the layer itself works as strong pinning centers of the vortices, resulting in that the diamagnetic effect may depend on the sites in a unit cell \cite{Tachiki}. Although in $\mathrm{Na_{0.35}CoO_{2} \cdot 1.3H_{2}O}$, no direct evidence of the intrinsic pinning was obtained, the Josephson vortex can be expected because of the strong two-dimensionality of this compound. Then the invariant behavior of Knight shift at about 7 T should be intrinsic, suggesting the spin triplet superconductivity.

If the change of the resonance fields below $T_{\mathrm{c}}$ at low fields shown in Fig. \ref{fig:spectraLF} is not attributed to the diamagnetic effect but due to the change of the intrinsic spin shift, the spin shift decreases with decreasing temperature. In this case, the spin triplet superconductivity with the superconducting $d$-vector perpendicular to the $c$-axis as well as the case of the spin singlet superconductivity can be considered. Recent theoretical study suggested that the hole-pocket near the K-point leads the ferromagnetic fluctuation, resulting in the spin triplet $p$- or $f$-wave superconductivity with the $d$-vector pointing perpendicular to the $c$-axis in the presence of the strong spin-orbit coupling \cite{Mochizuki,Yanase}. A part of the in-plane spin shifts perpendicular to the $d$-vector may be suppressed at $T$ = 0 K. The magnetic field induces the local density of states around the vortex cores, and then a part of the in-plane shift recovers. In this case, the key point is the Knight shift along the $c$-axis, $K_{z}$. In the preliminary results using the resonance center, which is due to the accidental alignment along the $c$-axis already mentioned, the temperature dependence of $K_{z}$ above $T_{\mathrm{c}}$ should be very small, suggesting the small contribution of the spin shift, and at least, $K_{z}$ does not change below $T_{\mathrm{c}}$. 

\subsection{$\mathrm{^{59}Co}$ nuclear spin-lattice relaxation time $T\mathrm{_{1}}$}
\begin{figure}[h]
 \begin{center}
 \includegraphics[width=1\linewidth]{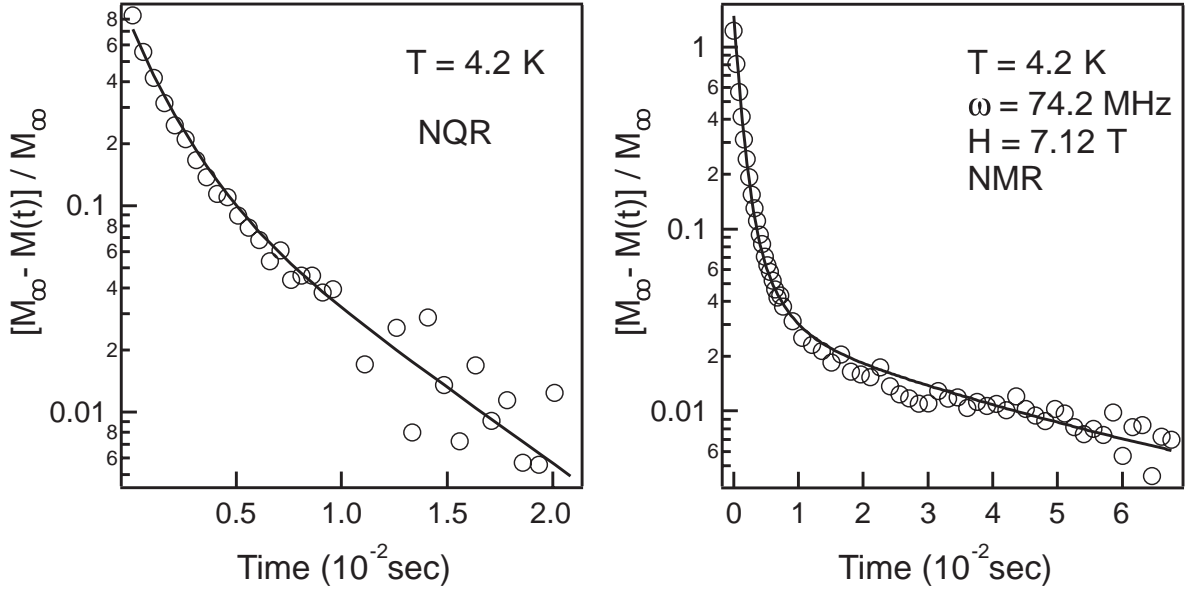}
 \end{center}
 \caption{Recovery curves at 4.2 K (a) for the highest frequency transition of $\mathrm{^{59}Co}$ NQR and (b) the center transition of the NMR taken at 74.2 MHz. The values of $1/T_1$ were estimated by the least-square fitting (see the text).}
 \label{fig:recovery}
 \end{figure}

\begin{figure}[h]
 \begin{center}
 \includegraphics[width=0.8\linewidth]{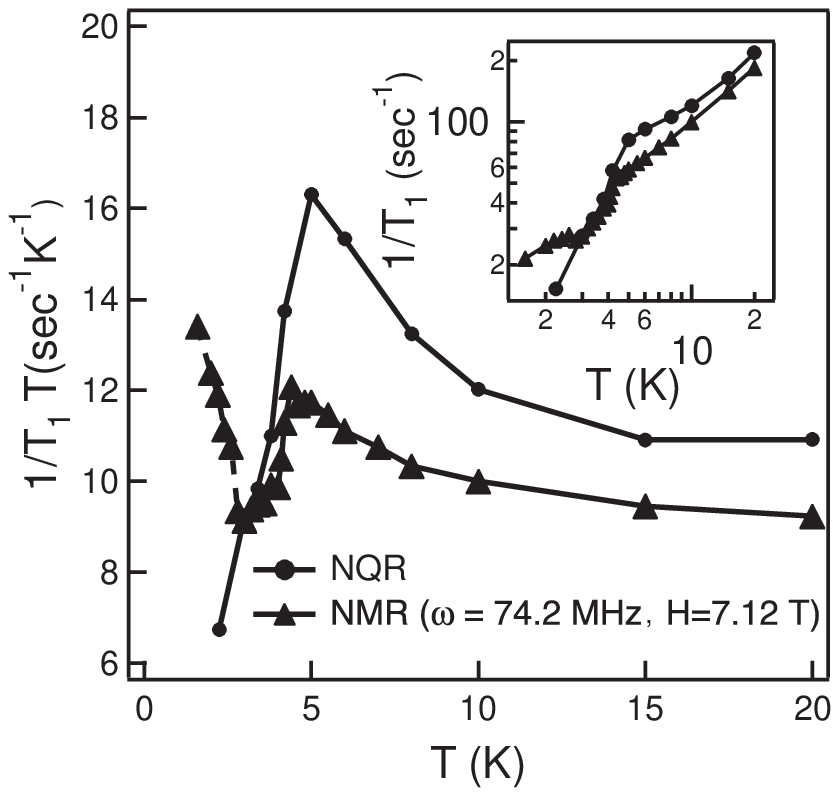}
 \end{center}
 \caption{\label{fig:T1inv}Temperature dependence of $1/T_1T$ for {\NaCo}. Filled circles are by the $^{59}$Co NQR data at zero external field and triangles are those of NMR at 7.12 T. The inset shows the temperature dependence of $1/T_1$. One can note that $1/T_1$ decreases proportional to $T^{3}$ just below $T\mathrm{_{c}}$ even under $H$ =7 T. }
 \end{figure}

\hspace*{10mm} We measured $1/T_{1}$ of $^{59}$Co under the central resonance field at frequencies of $H_2$ in order to obtain the information of low lying quasiparticle states below and above $T\mathrm{_{c}}$. 
In Fig. \ref{fig:recovery}, we show recovery curves at 4.2 K measured by NQR and NMR (at $H_2$) methods. 
The recovery curves for NQR signals at 4.2 K were analyzed by the same function as reported in Refs. \cite{Fujimoto} and \cite{Ishida4}. In the case of NMR, the observed recovery curves were found to consist of more than one relaxation component  maybe due to the overlap of satellite resonances in the powder spectrum. 
Thus, we should utilize the combination of the theoretical relaxation function for the central resonance and a slow single exponential component as follows, 
\begin{eqnarray}
\fl \frac{M(\infty)-M(t)}{M(\infty)}=
A\left( 
\frac{1}{84}e^{\frac{-t}{T_1}} + \frac{3}{44}e^{\frac{-6t}{T_1}} + 
\frac{75}{364}e^{\frac{-15t}{T_1}} + \frac{1225}{1716}e^{\frac{-28t}{T_1}}
\right) + 
Be^{\frac{-t}{T_1^{\mathrm{slow}}}},
\label{eq:recovery}
\end{eqnarray} 
where $A$, $B$, $T_{1}$ and $T_1^{\mathrm{slow}}$ are fitting parameters. 
The solid lines in Fig. \ref{fig:recovery} are the results of the least-square fitting using Eq. (\ref{eq:recovery}). 
The extrinsic slow component was very small, in which $B/(A+B)$ was found to be about 0.8 \%.
Figure \ref{fig:T1inv} shows the temperature dependence of $1/T_{1}T$ and the inset of the figure shows that of $1/T_{1}T$. 
In NQR measurement, $1/T_{1}$ rapidly decreases with temperature below $T\mathrm{_{c}}$ with $T^{3}$-dependence. 
This result is consistent with the data previously reported \cite{Fujimoto,Ishida4}.

Under the field of 7.12 T, the rapid decrease in $1/T_{1}T$ is also observed below 4.4 K, which indicates that $T\mathrm{_{c}}$ does not decrease markedly by the external field. 
We would like to note that $H\mathrm{_{c2}}$ under the magnetic field parallel to the $\mathrm{CoO_{2}}$ plane is quite large and $T\mathrm{_{c}}$ decreases very small (estimated about only 0.3 K smaller than that at zero field) at about 7 T. 
It is also noted that another enhancement of $1/T_{1}T$ below $T\mathrm{_{c}}$ which may suggest the presence of the unconventional vortex state because this behavior cannot be observed in the NQR measurement under zero magnetic field. 

The $T^{3}$ dependence of $1/T_1$ below $T\mathrm{_{c}}$ without any coherence peaks indicates the presence of line nodes on the Fermi surface. 
The recent study on the mother compound Na$_{x}\mathrm{CoO_{2}}$ revealed that the ferromagnetic spin fluctuations are dominant in the magnetism on the $\mathrm{CoO_{2}}$ plane \cite{Boothroyd}. 
For our hydrated \NaCo, one should note in Fig. \ref{fig:T1inv} that the enhancement of $1/T_1T$ in NQR just above $T\mathrm{_{c}}$ is suppressed at 7.12 T, which indicates that the ferromagnetic fluctuation is dominant in the same temperature range as also suggested by the enhancement of the magnetic susceptibility in Fig. \ref{fig:KTbelowTc}. This situation can be suitable for the occurrence of the spin-triplet superconductivity.

\section{Summary}
\hspace*{10mm} From the analysis of the temperature dependence of the field-swept NMR spectra, we estimated the spin component of $\mathrm{^{59}Co}$ Knight shift of {\NaCo}. 
In the normal state, the in-plane spin Knight shifts were found to be proportional to $\chi$ and to have the values of 0.3 and 0.1 $\%$ for $K_{x}$ and $K_{y}$ at least just above $T\mathrm{_{c}}$.
In the superconducting state, the intrinsic Knight shift does not change with temperature at the external field of about 7 T much smaller than $H\mathrm{_{c2}}$, although the Knight shift is not estimated correctly at lower fields because of lack of information of the precise diamagnetic effects.
We emphasize that the invariant behavior of the Knight shift in higher fields is consistent with that in $\mu$SR experiment \cite{higemoto}.
From these results, we conclude that the $p$- or $f$- wave pairing state with triplet spin symmetry may be most preferable in the superconducting state of {\NaCo}. 

\ack{This study was supported by a Grant-in-Aid on priority area
 'Novel Quantum Phenomena in Transition Metal Oxides',
 from Ministry of Education, Science, Sports and Culture (12046241, 16076210),
 and also supported by a Grant-in-Aid Scientific Research of 
 Japan Society for Promotion of Science
 (14654060, 14740382, 17750059).
}


\end{document}